\documentclass[aps,prd,twocolumn,showpacs,superscriptaddress,preprintnumbers]{revtex4-1}
\usepackage[colorlinks=true, pdfstartview=FitV, linkcolor=red,citecolor=blue, urlcolor=blue,
bookmarks=true, bookmarksnumbered=true, breaklinks]{hyperref}
\usepackage[dvipdfmx]{graphicx}
\usepackage{amsmath,amssymb,tabularx,mathrsfs}

\allowdisplaybreaks
\newcommand{\Slash}[1]{{\ooalign{\hfil/\hfil\crcr$#1$}}} 

\begin{document}

\title{$D$ meson mass increase by restoration of chiral symmetry in nuclear matter}

\author{Kei Suzuki}
\email{kei.suzuki@riken.jp}
\affiliation{Theoretical Research Division, Nishina Center, RIKEN, Wako, Saitama, 351-0198, Japan}
\author{Philipp Gubler}
\email{pgubler@riken.jp}
\affiliation{ECT$^*$, Villa Tambosi, 38123 Villazzano (Trento), Italy}
\author{Makoto Oka}
\email{oka@th.phys.titech.ac.jp}
\affiliation{Department of Physics, Tokyo Institute of Technology, Meguro, Tokyo, 152-8551, Japan}
\affiliation{Advanced Science Research Center, Japan Atomic Energy Agency, Tokai, Ibaraki, 319-1195, Japan}

\preprint{RIKEN-QHP-205}

\begin{abstract}
Spectral functions of the pseudoscalar $D$ meson in the nuclear medium are analyzed using QCD sum rules and the maximum entropy method.
This approach enables us to extract the spectral functions without any phenomenological assumption, and thus to visualize in-medium modification of the spectral functions directly.
It is found that 
the reduction of the chiral condensates of dimension 3 and 5 causes 
the masses of both $D^+$ and $D^-$ mesons to grow gradually at finite density. 
Additionally, we construct charge-conjugate-projected sum rules and find a $D^+$--$D^-$ mass splitting of about $-15$ MeV at nuclear saturation density. 
\end{abstract}
\maketitle

\section{Introduction} \label{Section_Introduction}
One of the most important problems in hadron physics is understanding the relation between chiral symmetry and hadron properties from Quantum chromodynamics (QCD) which describes the strong interaction.
Hadrons in nuclear matter are useful as probes of chiral symmetry at finite density.
For instance, $\rho$, $\omega$, and $\phi$ mesons in nuclear matter have been studied theoretically and experimentally (see Refs.~\cite{Hayano:2008vn, Leupold:2009kz} for reviews).
In the future, J-PARC as well as the compressed baryonic matter CBM) \cite{Friman:2011zz} and PANDA \cite{Lutz:2009ff} experiments by Facility for Antiproton and Ion Research (FAIR) at GSI are expected to investigate the properties of open ($D, \bar{D}$) and hidden ($J/\psi, \eta_c$) charmed mesons in hot and dense baryonic matter.

Medium modifications of pseudoscalar $D$ mesons in nuclear matter have been investigated in various theoretical studies. 
These can be classified into two approaches: theories based on hadron and those based on quark and gluon degrees of freedom.
The former, described by interactions between hadrons, includes self-consistent unitarized coupled-channel approaches with flavor $SU(3)$ \cite{Tolos:2004yg}, 
flavor $SU(4)$ symmetry, and a t-channel vector meson exchange (TVME) model \cite{Lutz:2005vx}, 
which have been further developed through an improved kernel and renormalization scheme \cite{Mizutani:2006vq,Tolos:2007vh}, improvement beyond 
zero range approximation \cite{JimenezTejero:2011fc}, and a $SU(8)$ spin-flavor symmetric model implementing 
heavy quark spin symmetry \cite{Tolos:2009nn}. 
There are furthermore results from a chiral $SU(3)$ model extended to $SU(4)$ \cite{Mishra:2003se,Mishra:2008cd,Kumar:2010gb,Kumar:2011ff}
and a pion exchange model between $\bar{D}$ and $N$ \cite{Yasui:2012rw}. 
The second approach includes the quark-meson coupling (QMC) model \cite{Tsushima:1998ru,Sibirtsev:1999js} and QCD sum rules \cite{Morath:1999cv,Hayashigaki:2000es,Hilger:2008jg,Azizi:2014bba,Wang:2015uya}.
Among all these, only QCD sum rules are directly based on QCD.

The QCD sum rule method \cite{Shifman:1978bx,Shifman:1978by} is known as a powerful tool to investigate the properties of hadrons from QCD.
It has also been used to study nuclear modifications of light meson systems such as $\rho$, $\omega$, and $\phi$ mesons \cite{Hatsuda:1991ez,Asakawa:1993pq,Asakawa:1993pm,Asakawa:1994tp,Hatsuda:1995dy,Jin:1995qg,Koike:1996ga,Klingl:1997kf,Lee:1997zta,Leupold:1997dg,Leupold:1998bt,Leupold:2001hj,Zschocke:2002mn,Kampfer:2002pj,Kampfer:2003sq,Zschocke:2004qa,Ruppert:2005id,Thomas:2005dc,Steinmueller:2006id,Kwon:2008vq,Hilger:2010cn,Gubler:2014pta,Gubler:2015uza,Gubler:2015yna}.
Recently, it has become possible to apply the maximum entropy method (MEM) to QCD sum rules \cite{Gubler:2010cf}, which allows us to extract the most probable form of the spectral function from the operator product expansion (OPE) of hadronic correlators without assuming any specific functional form, e.g., the ``pole + continuum'' ansatz.
This approach was shown to be successful in vacuum for the $\rho$ meson \cite{Gubler:2010cf} and the nucleon with positive \cite{Ohtani:2011zz} 
and negative parity \cite{Ohtani:2012ps}. 
Furthermore, it was used to investigate 
spectral modifications at finite temperature for charmonia \cite{Gubler:2011ua} and bottomonia \cite{Suzuki:2012ze} channels and at finite density for the $\phi$ meson \cite{Gubler:2014pta}.

Previous QCD sum rule studies of in-medium $D$ mesons have led to somewhat inconsistent results. 
In Ref.~\cite{Hayashigaki:2000es}, Hayashigaki calculated the OPE, including condensates up to dimension--4, $\langle \bar{q} q \rangle$, $\langle \frac{\alpha_s}{\pi} G^2 \rangle$, $\langle q^\dag i \overrightarrow{ D }_0 q \rangle$, and $\langle \frac{\alpha_s}{\pi} \left( \frac{(vG)^2}{v^2} - \frac{G^2}{4}\right) \rangle$, and analyzed the $D$ meson mass.
He found that the $D$ meson mass is shifted by $-50 \mathrm{MeV}$ at nuclear saturation density $\rho_0$.
Subsequently, Hilger {\it et al.} \cite{Hilger:2008jg} took further condensates up to dimension--5, $ \langle \bar{q} g \sigma G q \rangle$, and $q_0$-odd terms, $ \langle q^\dag q \rangle, \langle q^\dag \overrightarrow{ D }_0^2 q \rangle, \langle q^\dag g \sigma G q \rangle$ into account. 
As a result, an opposite mass shift of $+45$ MeV at $\rho_0$ was obtained, however, with a significant ambiguity from phenomenological density dependence of the threshold parameter. 
Recent new analyses \cite{Azizi:2014bba,Wang:2015uya} support the conclusions of \cite{Hayashigaki:2000es}. 
The results of the present paper, obtained by applying the MEM to the QCD sum rules, are independent of uncertainties from phenomenological functional forms and its threshold parameter.

Hilger {\it et al.} furthermore evaluated the $D^+$--$D^-$ mass splitting to be $-60$ MeV at $\rho_0$ \cite{Hilger:2008jg}. 
It is important to note that in contrast to the $q_0$-even terms, the $q_0$-odd terms violate the charge symmetry of the hadronic correlator and hence lead to a mass splitting 
of the $D^+$ and $D^-$ states. 
This charge-symmetry breaking comes from the asymmetry of the nuclear medium, which consists only of nucleons (or only quarks) and not of antinucleons (or antiquarks). 
The properties of the $D$ ($D^+ = c\bar{d}$ and $D^0 = c\bar{u}$) and $\bar{D}$ mesons ($D^- = \bar{c}d$ and $\bar{D}^0 = \bar{c}u$) can therefore be different at finite density.
To improve the analysis of the $D^+$--$D^-$ mass splitting, we propose in this work the charge conjugate projection as a novel approach.

This paper is organized as follows.
In Sec.~\ref{Section_Formalism}, we present QCD sum rules of the $D$ meson in nuclear matter. 
In Sec.~\ref{Section_Results}, the results of our QCD sum rules and MEM analyses are reported and their physical interpretation is given.
In Sec.~\ref{Section_Comments}, we compare them with the previous QCD sum rule analyses.
Section~\ref{Section_Conclusion and Outlook} is devoted to the conclusion and outlook.

\section{Formalism} \label{Section_Formalism}
We start by defining the time-ordered hadronic current-current correlation function:
\begin{equation}
\Pi^{\, J}(q) = i \int d^4x e^{i q \cdot x} \langle T[j^{\, J}(x) j^{\, J \dag}(0)] \rangle , \label{correlation function}
\end{equation}
where $J$ stands for the channels $D^+(c\bar{d})$, $D^-(\bar{c}d)$, $D^0(c\bar{u})$, or $\bar{D}^0(\bar{c}u)$.
In this work, we assume the chiral limit ($m_u = m_d =0$) and isospin symmetry ($\langle \bar{u}u \rangle = \langle \bar{d}d \rangle$), so that $u$ and $d$ quarks are not distinguished.
We thus only need to 
examine $j^{\, D^+}= i \bar{d} \gamma_5 c$ and $ j^{\, D^-} = i \bar{c} \gamma_5 d$ as possible pseudoscalar currents.

$\Pi^{\, J}(q)$ satisfies the dispersion relation in momentum space given as
\begin{eqnarray}
\Pi^{\, J}(q^2) &=& \frac{1}{\pi} \int_0^\infty ds \frac{\mathrm{Im} \Pi^{\, J}(s+i \epsilon)}{ s-q^2} \nonumber\\
&\equiv&\int_0^\infty  ds\frac{\rho^{\, J} (s)}{s-q^2}. \label{dispertion_relation}
\end{eqnarray}
We calculate the left-hand side using the operator product expansion (OPE) in the large Euclidean momentum ($Q^2=-q^2<0$) region, 
where QCD can be treated perturbatively thanks to asymptotic freedom. 
As a next step, one usually 
deforms the kernel by transforming both sides of Eq.~(\ref{dispertion_relation}) by Borel or Gaussian transformations.
In this work, we employ the Gaussian sum rule \cite{Bertlmann:1984ih,Orlandini:2000nv},  
which has a number of advantages over the conventional Borel sum rule. 
The Gaussian sum rule has two controllable parameters, $\hat{s}$ and $\tau$, while its Borel counterpart allows only one (the Borel mass). 
The transformed kernel of the integral equation can therefore describe 
various shapes, depending on both $\hat{s}$ and $\tau$, 
so that more patterns of weight functions can be used to extract spectral functions from the MEM analysis.
With the Gaussian sum rule it is furthermore possible to decrease some statistical errors originating from MEM. 

\subsection{OPE in vacuum}
After the Gaussian transformation, the dispersion relation becomes 
\begin{equation}
G(\hat{s}, \tau) = \frac{1}{\sqrt{4\pi \tau}} \int_0^\infty d\omega \, \omega e^{-\frac{(\omega^2 - \hat{s})^2}{4\tau} } \rho (\omega^2), \label{Gaussian sum rule}
\end{equation}
where $\omega$ denotes the energy ($\omega^2=s$) and $\hat{s}$ and $\tau$ are the parameters of the Gaussian transformation.
The new kernel enhances 
an energy region of the spectral function around the position $\hat{s}$. 
By tuning $\hat{s}$ and $\tau$, one can focus on the lowest peak and suppress the higher energy structures such as excited states and continuum by the tail of the Gaussian.

The OPE including up to the dimension--5 condensates was calculated for the pseudoscalar $D$ meson ($J^\pi = 0^-$) in Ref.~\cite{Aliev:1983ra}. 
Its Gaussian-transform is given as
\begin{eqnarray}
&& G(\hat{s}, \tau) = \frac{1}{\sqrt{4\pi \tau}} \frac{1}{\pi} \int_{m_h^2}^\infty \frac{ds}{2} e^{  -\frac{(s - \hat{s})^2}{4\tau}  } \mathrm{Im} \Pi^{ \mathrm{pert}} (s)  \nonumber\\
&& +  \frac{1}{\sqrt{4\pi \tau}} e^{  -\frac{(m_h^2 - \hat{s})^2}{4\tau}  } \left[ -m_h \langle \bar{q} q \rangle + \frac{1}{12} \langle \frac{\alpha}{\pi} G^2 \rangle  \right. \nonumber\\
&& \left. - \frac{1}{2} \left( \frac{3m_h^2 - 2\hat{s}}{ 4\tau} - \frac{ 2(m_h^2 - \hat{s})^2 m_h^2 }{ (4 \tau )^2 } \right) m_h \langle \bar{q} g \sigma G q \rangle  \right], \label{OPE_vacuum}
\end{eqnarray}
where $m_h$ is a general heavy quark mass, which will be set to the charm quark mass ($m_c$) for the most part of this work.
In Sec.~\ref{Subsection_mh_dep}, $m_h$ will, however, be treated as a free parameter to investigate the heavy quark mass dependence of the sum rules.
The perturbative term, $ \mathrm{Im} \, \Pi^{ \mathrm{pert}} (s)$, including first-order $\alpha_s$ corrections, is given by
\begin{equation}
\mathrm{Im} \, \Pi^{ \mathrm{pert}} (s) = \frac{3}{8\pi} s \left( 1- \frac{m_h^2}{s}\right)^2 \times \left( 1 + \frac{4}{3} \frac{\alpha_s}{\pi} R_0(m_h^2/s) \right),
\end{equation}
where
\begin{eqnarray}
R_0( x) &=& \frac{9}{4} + 2Li_2(x)  +\ln x \ln (1-x) -\frac{3}{2} \ln \frac{1-x}{x} \nonumber\\
&& -\ln (1-x) + x \ln \frac{1-x}{x} -\frac{x}{1-x} \ln x. \label{pert_term}
\end{eqnarray}

\subsection{OPE in nuclear medium}

In this work, we choose our reference frame as the rest frame of the nuclear medium, and we set the spatial momentum of the meson to zero: $q=(q_0, \bf{0})$.
In the vacuum, the OPE depends only on $q^2$ because of Lorentz invariance while, at finite density, we have to take into account the terms of odd powers of $q_0$.
The correlator is hence separated into $q_0$-even and $q_0$-odd parts:
\begin{equation}
\Pi^{\, J}(q_0) = \Pi^{\, \rm{even}}(q_0^2) + q_0 \Pi^{\, \rm{odd}}(q_0^2). \label{even-odd}
\end{equation}
As long as we consider a system at low enough density (such as nuclear matter), the Wilson coefficients can be assumed to have no density dependence and it suffices 
to include density dependencies of the condensates. 
The separated $q_0$-even and $q_0$-odd parts of the $D$ meson OPE at finite density were derived in Ref.~\cite{Zschocke:2011aa}.
Additionally, the OPE including dimension--6 condensates in medium was estimated in Ref.~\cite{Buchheim:2014rpa}. 

\subsection{Charge conjugate projection}
\begin{figure}[t!]
     \begin{center}
    \begin{minipage}[t]{1.0\columnwidth}
        \begin{center}
            \includegraphics[clip, width=1.0\columnwidth]{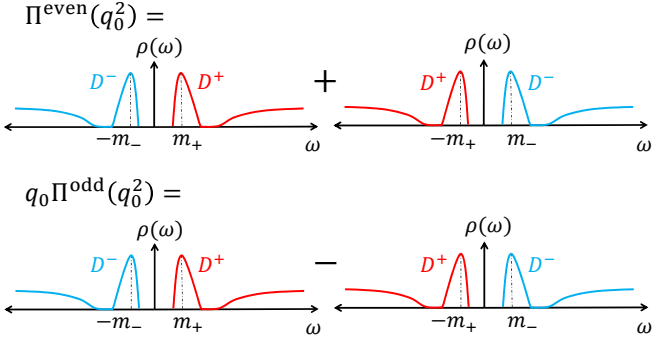}
        \end{center}
    \end{minipage}
    \end{center}
     \caption{Schematic picture of spectral function contributions in $\Pi^{\, \rm{even}}(q_0^2)$ and $q_0 \Pi^{\, \rm{odd}}(q_0^2)$ of the $D^+$ correlator. 
The factor $1/2$ on the right hand side is omitted for simplicity. 
Spectral functions of the old-fashioned correlator 
include only spectra in the positive energy region.}
     \label{charge_conjugate_projection}
\end{figure}

$\Pi^{\, \rm{even}} (q_0^2)$ and $\Pi^{\, \rm{odd}} (q_0^2)$ in momentum space contain information from $D^+$ and $D^-$ spectra in 
both positive and negative energy regions. 
Namely, each term of the correlator $\Pi^{\, D^+}(q_0)$ of Eq.~(\ref{even-odd}), can be rewritten as

\begin{eqnarray}
\Pi^{\, \rm{even}}(q_0^2) &=& \tfrac{1}{2} \left[ \Pi^{\, +}(q_0) + \Pi^{\, -}(q_0) \right], \\
q_0 \Pi^{\, \rm{odd}}(q_0^2) &=& \tfrac{1}{2}  \left[ \Pi^{\, +}(q_0) - \Pi^{\, -}(q_0) \right] ,
\end{eqnarray}
where $\Pi^{\, +}$($\Pi^{\, -}$) corresponds to the $D^+$($D^-$) spectrum for positive energy and the $D^-$($D^+$) spectrum for negative energy (see Fig.~\ref{charge_conjugate_projection}).
For $\Pi^{\, D^-}(q_0)$, the situation is reversed. 
To separate $D^+$ and $D^-$ from $\Pi^{\, J}(q_0)$, we will formulate below the {\it charge-conjugate-projected sum rule}, which is analogous to 
the parity projection for baryon sum rules \cite{Jido:1996ia, Kondo:2005ur, Ohtani:2012ps}.

In this approach, we define the {\it old-fashioned} correlator in the rest frame ($q=(q_0, \bf{0})$):
\begin{equation}
\left[ \Pi^{\, J}(q_0)\right]^{ \rm{old}} = i \int d^4x e^{i q_0 x_0} \theta(x_0) \langle T[j^{\, J}(x) j^{\, J \dag}(0)] \rangle,
\end{equation}
where $\theta(x)$ is the Heaviside step function which removes the negative energy contibution from the correlator. 
Using the $q_0$-even and $q_0$-odd parts, the new OPE of the charge-conjugate-projected sum rules is given as
\begin{equation}
\Pi^{\, D^\pm}_{ \rm{OPE}} \equiv  \left[ \Pi^{\, \pm}(q_0) \right]^{ \rm{old}} = \left[ \Pi^{\, \rm{even}}(q_0^2) \pm q_0 \Pi^{\, \rm{odd}}(q_0^2) \right]^{ \rm{old}}. \label{OPE}
\end{equation}
Analyticity of the correlation functions connects the projected spectral functions to the imaginary part of the projected OPE.
Multiplying the Gaussian kernel 
$W(q_0, \hat{s}, \tau) =  \frac{q_0}{\sqrt{4\pi \tau}} \exp[  -\frac{(q_0^2 - \hat{s})^2}{4\tau}  ] $ as a weight function, 
we obtain the following integral sum rules:
\begin{eqnarray}
&& \int_{-\infty}^\infty dq_0 \, \frac{1}{\pi} \mathrm{Im} \, \Pi^{\, D^\pm}_{ \rm{OPE}} \, W(q_0, \hat{s}, \tau) \nonumber \\ 
&& \hspace{50pt} = \tilde{G}^\mathrm{even} (\hat{s}, \tau) \pm \tilde{G}^\mathrm{odd} (\hat{s}, \tau)  \nonumber \\
&& \hspace{50pt} = \int_0^\infty d\omega \, \rho^\pm (\omega) \, W(\omega, \hat{s}, \tau). \label{sum rule}
\end{eqnarray}
Here, Gaussian-transformed $q_0$-even and $q_0$-odd parts are defined as
\begin{eqnarray}
&& \tilde{G}^\mathrm{even} (\hat{s}, \tau) = \int_0^\infty dq_0 \, W(q_0, \hat{s}, \tau) \frac{1}{\pi} \mathrm{Im} \left[ \Pi^{\, \rm{even}}(q_0^2)  \right]^{ \rm{old}} \nonumber \\
&& \tilde{G}^\mathrm{odd} (\hat{s}, \tau) = \int_0^\infty dq_0 \, W(q_0, \hat{s}, \tau) \frac{1}{\pi} \mathrm{Im} \left[ q_0 \Pi^{\, \rm{odd}}(q_0^2) \right]^{ \rm{old}}. \nonumber \\ \label{Gauss_trans}
\end{eqnarray}
With this definition, we reach the final form of the charge-conjugate-projected OPE for $D^+$ and $D^-$ mesons in nuclear medium:
\begin{widetext}
\begin{eqnarray}
\tilde{G}^\mathrm{even}(\hat{s}, \tau) &=& \frac{1}{2\sqrt{4\pi \tau}} \frac{1}{\pi} \int_{m_h^2}^\infty \frac{ds}{2} e^{  -\frac{(s - \hat{s})^2}{4\tau}  } \mathrm{Im} \Pi^{ \mathrm{pert}} (s)  \nonumber\\
&& + \frac{1}{2\sqrt{4\pi \tau}} e^{  -\frac{(m_h^2 - \hat{s})^2}{4\tau}  } \left[ -m_h \langle \bar{q} q \rangle + \frac{1}{12} \langle \frac{\alpha}{\pi} G^2 \rangle
 - \frac{1}{2} \left( \frac{3m_h^2 - 2\hat{s}}{ 4\tau} - \frac{ 2(m_h^2 - \hat{s})^2 m_h^2 }{ (4 \tau )^2 } \right) m_h \langle \bar{q} g \sigma G q \rangle  \right. \nonumber\\
&& \hspace{58pt} + \left\{ \frac{1}{9}-\frac{5m_h^2}{36\tau}(m_h^2 - \hat{s}) +\left( -\frac{1}{3} +\frac{m_h^2(m_h^2 - \hat{s})}{6\tau} \right) \ln{\frac{\mu^2}{4m_h^2}} \right\} \langle \frac{\alpha_s}{\pi} \left( \frac{(vG)^2}{v^2} - \frac{G^2}{4}\right) \rangle \nonumber\\
&& \hspace{58pt} - 2 \left( 1 - \frac{ (m_h^2 - \hat{s} ) m_h^2}{2 \tau} \right) \langle q^\dag i \overrightarrow{ D }_0 q \rangle  \nonumber\\
&& \left. \hspace{58pt} - 4 \left( \frac{3m_h^2 - 2\hat{s}}{ 4\tau} - \frac{ 2(m_h^2 - \hat{s})^2 m_h^2 }{ (4 \tau )^2 } \right) m_h \left[ \langle \bar{q} \overrightarrow{ D }_0^2 q \rangle - \langle \frac{1}{8} \bar{q}  g \sigma G q \rangle \right] \right] \nonumber\\
&& \hspace{-70pt} + \frac{1}{2\sqrt{4\pi \tau}} \int_0^\infty dy e^{- \frac{[m_h^2(1+y)^2 - \hat{s}]^2}{4\tau}} \left\{ -\frac{1}{3} \frac{(1+y)^2}{(2+y)^2} -\frac{\ln{y}}{3\tau^2} \left[ m_h^8 (1+y)^7 -2m_h^6 \hat{s}(1+y)^5 +m_h^4 (1+y)^3 (\hat{s}^2-(6+y)\tau) \right.\right. \nonumber\\
&& \hspace{+73pt} \left. \left. +m_h^2\hat{s}(4+5y+y^2) \tau + \tau^2 \right] \right\} \times \langle \frac{\alpha_s}{\pi} \left( \frac{(vG)^2}{v^2} - \frac{G^2}{4} \right) \rangle, \label{even_term}
\end{eqnarray}
\begin{eqnarray}
\tilde{G}^\mathrm{odd}(\hat{s}, \tau) = \frac{1}{2\sqrt{4\pi \tau}} e^{  -\frac{(m_h^2 - \hat{s})^2}{4\tau}  } && \left[ m_h \langle q^\dag q \rangle  +4 \left( - \frac{3}{8m_h} + \frac{(4m_h^2 - 3\hat{s})m_h}{ 4\tau} - \frac{ 2(m_h^2 - \hat{s})^2 m_h^3 }{ (4 \tau )^2 } \right) \langle q^\dag \overrightarrow{ D }_0^2 q \rangle \right. \nonumber\\
&& \hspace{10pt} \left.  - \left( -\frac{1}{2m_h} + \frac{(m_h^2 - \hat{s}) m_h }{2\tau} \right) \langle q^\dag g \sigma G q \rangle \right]. \label{odd_term}
\end{eqnarray}
\end{widetext}
We note that perturbative, $\langle \bar{q} q \rangle$, $\langle \frac{\alpha}{\pi} G^2 \rangle$ and $\langle \bar{q} g \sigma G q \rangle$ terms in Eq.~(\ref{even_term}) 
agree with the OPE in vacuum of Eq.~(\ref{OPE_vacuum}) (times $1/2$). 
This factor can be understood from the fact that 
Eq.~(\ref{OPE_vacuum}) includes spectra from both $D^+$ and $D^-$, while they are separated in Eq.~(\ref{even_term}).

To extract the spectral functions $\rho^\pm (\omega)$ for $D^+$ and $D^-$ mesons from the sum rules of Eq.~(\ref{sum rule}), we employ the MEM \cite{Gubler:2010cf}.
The procedure of the MEM for Gaussian sum rules is summarized in Appendix~\ref{App_MEM}.

\section{Results} \label{Section_Results}

\begin{table*}
  \begin{center}
  \begin{tabular}{l|c|c}
\hline \hline
Condensates                                           & Vacuum value ($\mu=m_c$) & Density dependence ($\mu=1\mathrm{GeV}$) \\
\hline
$ \langle \bar{q} q \rangle$                          & $(-0.2685 (12)(14) \text{ GeV})^3$ \cite{Borsanyi:2012zv} & $(\sigma_{\pi N}/(m_u+m_d)) \, \rho$ \\
$ \langle \frac{\alpha}{\pi} G^2 \rangle $            & $(0.33 \pm 0.04 \text{ GeV})^4$    & $(- 0.65 \pm 0.15 \mathrm{GeV}) \, \rho$ \cite{Jin:1992id} \\
$ \langle \bar{q} g \sigma G q \rangle $              & $ (0.66 \pm 0.17 \mathrm{GeV}^2) \langle \bar{q} q \rangle$ & $ (3 \pm 1 \mathrm{GeV}^2) \, \rho$ \cite{Jin:1992id} \\
$ \langle q^\dag q \rangle $                          & $0$                      & $1.5 \, \rho$ \\
$ \langle \frac{\alpha_s}{\pi} \left( \frac{(vG)^2}{v^2} - \frac{G^2}{4} \right) \rangle $ & $0$ & $(- 0.042 \pm 0.017 \mathrm{GeV}) \, \rho$ \\
$ \langle q^\dag g \sigma G q \rangle $               & $0$                      & $(0.33 \mathrm{GeV}^2) \, \rho$ \cite{Braun:1986ty,Jin:1992id,Hilger:2008jg} \\
$ \langle q^\dag i \overrightarrow{ D }_0 q \rangle$  & $0$                      & $(0.218 \pm 0.021 \mathrm{GeV}) \, \rho$ \\
$ \langle \bar{q} \overrightarrow{ D }_0^2 q \rangle - \langle \frac{1}{8} \bar{q}  g \sigma G q \rangle $ & $0$ & $(-0.011 \pm 0.031 \mathrm{GeV}^2) \, \rho$ \\
$ \langle q^\dag \overrightarrow{ D }_0^2 q \rangle $ & $0$                      & $(-0.033 \pm 0.004 \mathrm{GeV}^2) \, \rho + \langle \frac{1}{12}q^\dag g \sigma G q \rangle$ \\
\hline \hline
    \end{tabular}
    \end{center}
    \caption{Numerical values of input parameters and those error bars in this work.
$\rho$ is the baryon number density.
Renormalization scale for condensates in vacuum is $\mu=m_c$.
In-medium condensates are shown values at $\mu=1\mathrm{GeV}$, where density dependence of $ \langle \bar{q} q \rangle$ and $ \langle \bar{q} g \sigma G q \rangle $ is run to $\mu=m_c$ in our numerical analyses.}
    \label{paramter_list}
\end{table*}

\subsection{Spectral functions in vacuum}
To extract the spectral function of the vacuum $D$ meson from the sum rule of Eq.~(\ref{sum rule}), we use the charm quark pole mass, $m_c (\mu=m_c) =1.67 \pm 0.07\,\mathrm{GeV}$ \cite{Agashe:2014kda}, and the strong coupling constant, $\alpha_s (\mu=m_c) = 0.337$ with $\Lambda_\mathrm{QCD}=0.296 \pm 0.013\,\mathrm{GeV}$ and the number of active flavors $N_f=4$ \cite{Bethke:2009jm}.
The used values of the condensates are shown in Table~\ref{paramter_list}. 
The error bars of these parameters are important because they are taken into account as uncertainties in the MEM analyses.

Next, we have to choose a range (so-called window) for the Gaussian parameters $\hat{s}$ and $\tau$, for which the OPE shows sufficient convergence. 
At both zero and non-zero densities, the dimension--3 $\langle \bar{q} q \rangle$ and dimension--5 $ \langle \bar{q} g \sigma G q \rangle $ terms 
mostly dominate the OPE. 
Additionally, contributions of dimension--6 quark condensates can be expected to be very small \cite{Buchheim:2014rpa} 
(for a short discussion of this point, see also Sec.~\ref{Subsection_dim6} and Appendix~\ref{App_dimension6} of this paper). 
We therefore, in this work, impose the criterion that the absolute value of the $ \langle \bar{q} g \sigma G q \rangle $ term should be less than $30 \%$ of the total OPE.
As long as this condition is satisfied, $\hat{s}$ and $\tau$ can be chosen arbitrarily. 
As a result, we use $1.67<\hat{s}<3.21 \, \mathrm{GeV}^2$ and $0.50<\tau<0.62 \, \mathrm{GeV}^4$. 

\begin{figure}[t!]
     \begin{center}
    \begin{minipage}[t]{1.0\columnwidth}
        \begin{center}
            \includegraphics[clip, width=1.0\columnwidth]{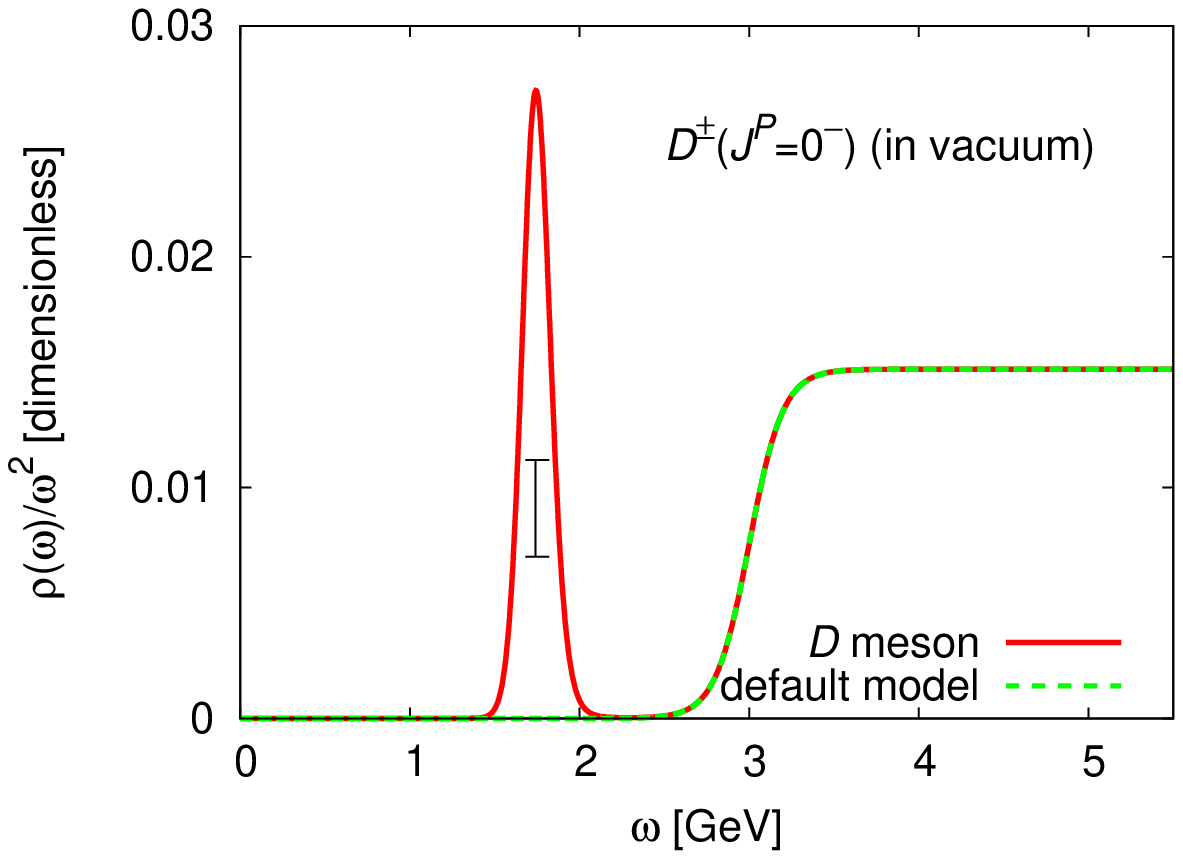}
        \end{center}
    \end{minipage}
    \end{center}
     \caption{Spectral function extracted with MEM from the $D^\pm$ meson sum rule in vacuum.
The definition of the error bar at the peak is given in Ref.~\cite{Gubler:2010cf}.}
     \label{SPFvacuum}
\end{figure}

Finally, we have to choose an input default model in the MEM analysis.
First, let us note that we will apply the MEM to the dimensionless function $\rho(\omega)/\omega^2$ in the present calculation. 
In this work, the function of the default model rises from nearly zero at a low energy to the value of the perturbative term of Eq.~(\ref{pert_term}) (divided by $s=\omega^2$) at high energy.
The concrete form of this function is shown in Fig.~\ref{SPFvacuum} as the green dashed line.

\begin{figure}[t!]
     \begin{center}
    \begin{minipage}[t]{1.0\columnwidth}
        \begin{center}
            \includegraphics[clip, width=1.0\columnwidth]{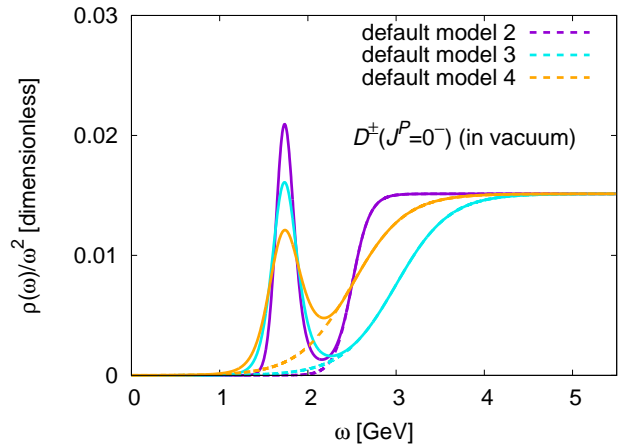}
        \end{center}
    \end{minipage}
    \end{center}
     \caption{Default model dependence of spectral function extracted with MEM from the $D^\pm$ meson sum rule in vacuum.
Solid and dashed lines denote spectral functions extracted with MEM and input default models, respectively.}
     \label{SPF_DM}
\end{figure}

The extracted $D$ meson spectral function in vacuum is shown as the red solid line in Fig.~\ref{SPFvacuum}. 
In vacuum, the $D^+$ and $D^-$ spectra are completely degenerate.
The first peak position is found at $1.74\mathrm{GeV}$, which has a systematic error which is typical ($\sim 10\%$) for QCD sum rule analyses.
The value is thus consistent with the experimental value of the $D$ meson ground-state mass of $1.87 \mathrm{GeV}$ \cite{Agashe:2014kda}.
In this work, we are only interested in relative mass shifts, for which the uncertainties of the absolute mass value largely cancel out.
The large systematic error for the absolute masses therefore do not prohibit the extraction of the small mass shifts to be discussed in this work.

Furthermore, the vertical error bar suggests that this peak is statistically significant.
Here, the central value of the error bar, denoted as $\langle \rho \rangle$, represents the value of $\rho(\omega)$ averaged over the energy range $\omega = 1.47-2.10 \mathrm{GeV}$ which corresponds to the overall range of the peak.
The upper and lower horizontal lines correspond to $\langle \rho \rangle \pm \langle \delta \rho \rangle$, where the error $\langle \delta \rho \rangle$ is determined by the MEM analysis \cite{Gubler:2010cf}.

To check the default model dependence of the spectral function, we obtained spectral functions with some functional forms, which is shown in Fig.~\ref{SPF_DM}.
From this figure, we find that although our results depend on the choice of the default model, the $D$ meson peak is always reproduced and its mass is almost independent of the default models.

\subsection{Spectral functions in nuclear medium}

Density dependencies of the vacuum condensates have been discussed in the past (see e.g., Refs.~\cite{Cohen:1991nk,Jin:1992id,Cohen:1994wm}).
For the $D$ meson system, the density dependence of the OPE is dominated by that 
of the $ \langle \bar{q} q \rangle$, $ \langle \bar{q} g \sigma G q \rangle $, and $ \langle q^\dag q \rangle $ terms 
and our final results are sensitive to the values of the corresponding parameters. 
The reduction of the quark condensate $\langle \bar{q} q \rangle$ at finite density is to leading order in $\rho$ governed by the $\pi N$ sigma term 
and the light quark masses, which we fix to $\sigma_{\pi N} =45 \pm 15 \mathrm{MeV}$ and $m_u+m_d=9 \pm 1 \mathrm{MeV}$ \cite{Agashe:2014kda} at $\mu=1 \mathrm{GeV}$. 
The behavior of the mixed condensate $ \langle \bar{q} g \sigma G q \rangle $ at finite density is much less well determined. 
Here, we follow the QCD sum rule literature and assume that its density dependence is proportional to the one of the quark condensate \cite{Jin:1992id}. 
$ \langle q^\dag q \rangle $ on the other hand is nothing but the expectation value of the quark density operator and its relation 
to the Baryon number density is therefore exact. 

\begin{figure}[t!]
    \begin{minipage}[t]{1.0\columnwidth}
        \begin{center}
            \includegraphics[clip, width=1.0\columnwidth]{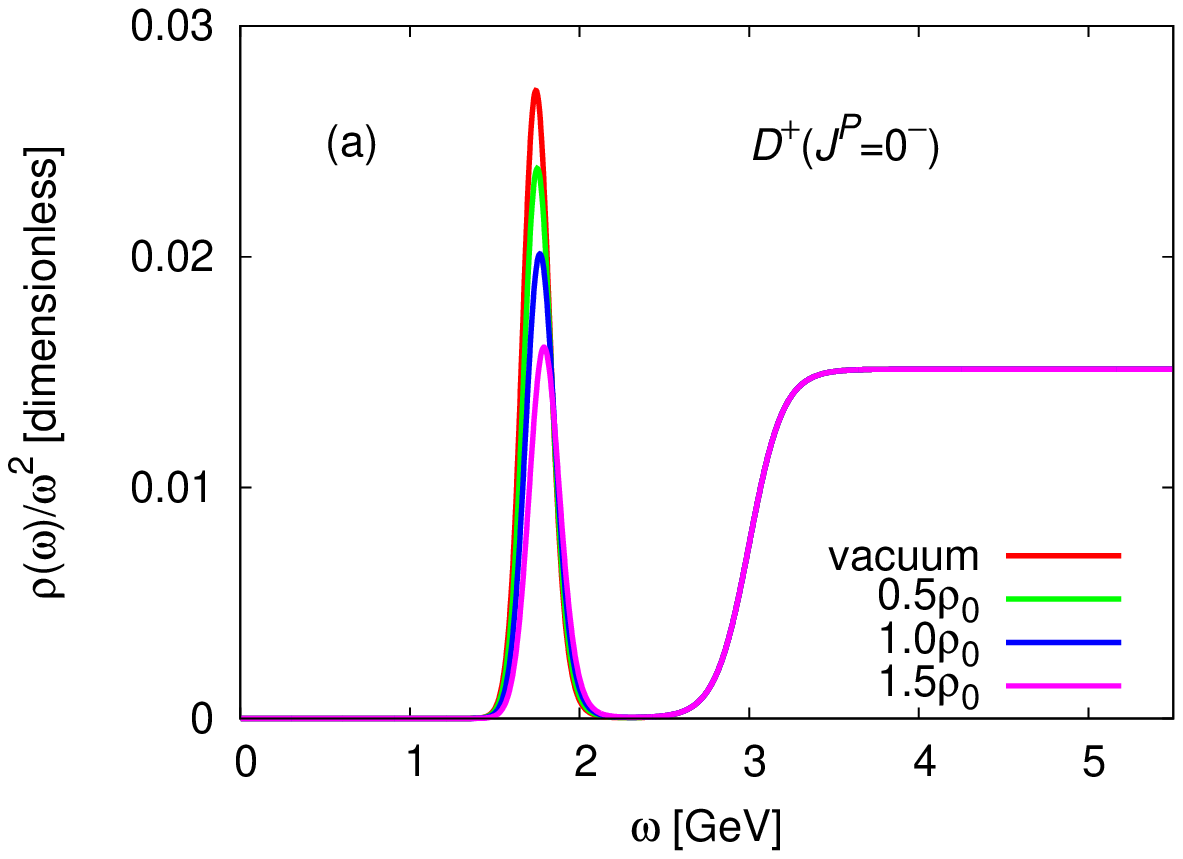}
        \end{center}
    \end{minipage}
    \begin{minipage}[t]{1.0\columnwidth}
        \begin{center}
            \includegraphics[clip, width=1.0\columnwidth]{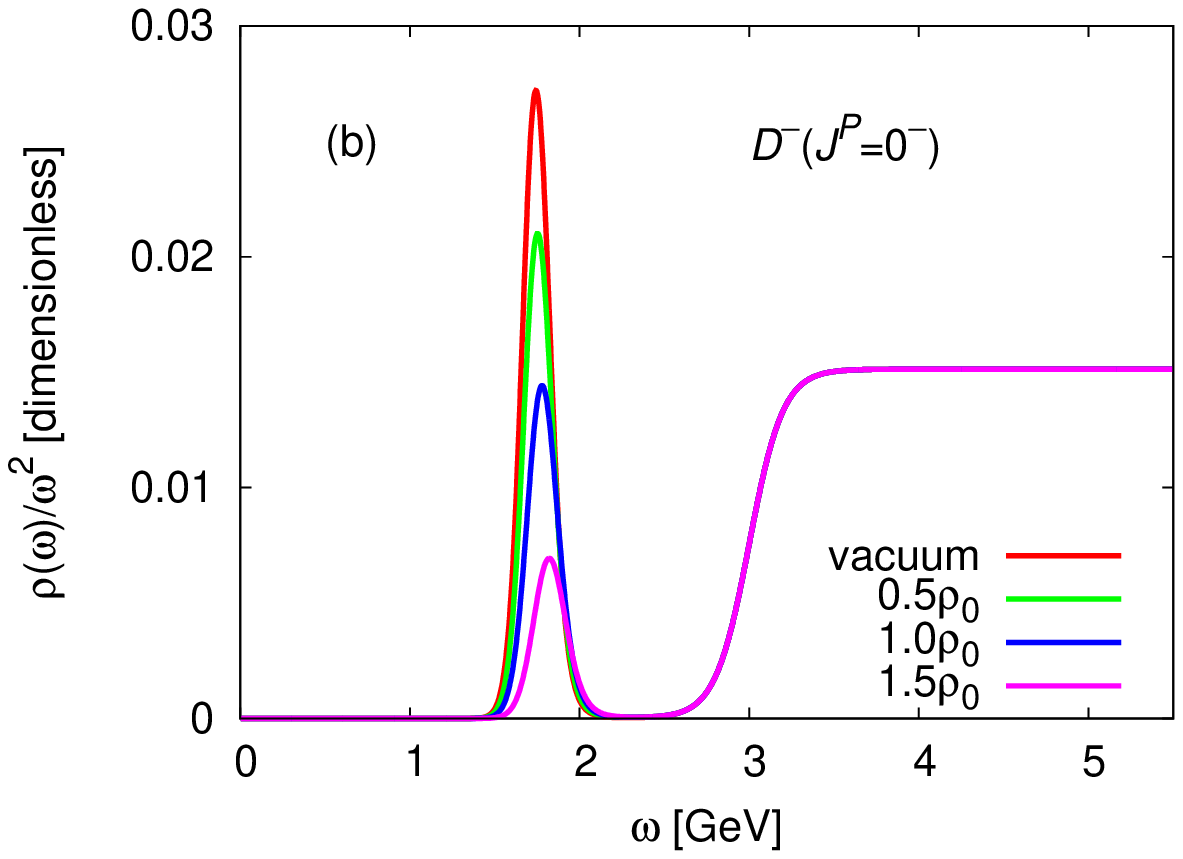}
        \end{center}
    \end{minipage}
    \caption{Spectral functions extracted with MEM from $D^\pm$ meson sum rules in nuclear matter.
(a) $D^+$ meson. (b) $D^-$ meson.
$\rho_0 = 0.0013 [\mathrm{GeV^3}]$ is the nuclear saturation density.}
    \label{SPFdensity}
\end{figure}

Next, let us discuss the nonscalar condensates, that show up only at finite density. 
As pointed out in Refs.~\cite{Jin:1992id,Hilger:2008jg}, not even the sign of $ \langle q^\dag g \sigma G q \rangle $ is known with certainty. 
Here, we choose the positive sign according to the reasons given in Ref.~\cite{Hilger:2008jg}.
Furthermore, to update the values of the derivative condensates, we apply the relations between nucleon matrix elements and 
moments of parton distributions as given in Ref.~\cite{Jin:1992id}: $ \langle  N| \frac{\alpha_s}{\pi} \left( \frac{(vG)^2}{v^2} - \frac{G^2}{4} \right) |N \rangle = -\frac{3}{4\pi}  M_N \alpha_s (\mu^2) A_2^g (\mu^2)$, $ \langle N|q^\dag i \overrightarrow{ D }_0 q|N \rangle = \frac{3}{8} M_N A_2^q (\mu^2)$, $ \langle N|\bar{q} \overrightarrow{ D }_0^2 q |N \rangle - \langle N| \frac{1}{8} \bar{q}  g \sigma G q |N \rangle = -\frac{3}{4} M_N^2 e_2^q (\mu^2)$, and $ \langle  N|q^\dag \overrightarrow{ D }_0^2 q|N \rangle - \langle N| \frac{1}{12} q^\dag g \sigma G q |N \rangle = -\frac{1}{4} M_N^2 A_3^q (\mu^2)$, where $M_N=0.939 \mathrm{GeV}$ is the nucleon mass.
The values of $A_2^g =0.359 \pm 0.146$, $A_2^q =0.62 \pm 0.06$ and $A_3^q=0.15\pm0.02$ are calculated by numerically integrating the parton distribution functions given in Ref.~\cite{Martin:2009iq}.
Also, $e_2^q =0.017 \pm 0.047$ is extracted from the recent experimental data of Ref.~\cite{Courtoy:2014ixa}, following the methods explained in Ref.~\cite{Gubler:2015uza}.
These quantities are averaged over $u$ and $d$ quarks and are given at a renormalization scale of about $1\mathrm{GeV}$.

\begin{figure}[t!]
     \begin{center}
    \begin{minipage}[t]{1.0\columnwidth}
        \begin{center}
            \includegraphics[clip, width=1.0\columnwidth]{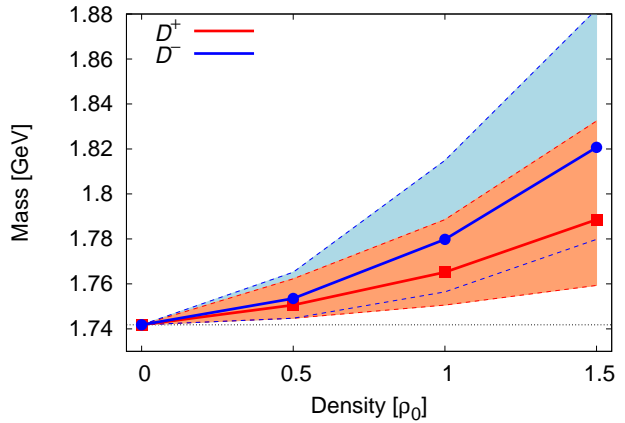}
        \end{center}
    \end{minipage}
    \end{center}
     \caption{Density dependence of $D^\pm$ meson peak positions. 
Dashed lines and shaded areas correspond to errors from uncertainties of in-medium condensates.
}
     \label{mass shift}
\end{figure}

{\it Discussion 1: Mass increase from chiral symmetry}.
Our results of the $D^\pm$ meson spectral functions at finite density are shown in Fig.~\ref{SPFdensity}.
For both $D^+$ and $D^-$, the peak residues gradually decrease as the density increases while the peak positions are shifted to higher energies.
The density dependencies of the peak positions are shown in Fig.~\ref{mass shift}.
Both $D^+$ and $D^-$ show positive energy shifts. 
As shall be discussed in more detail in the next section, 
we find that the main source of the mass increase is the density dependence of the chiral condensate.
Namely, the mass enhancements in the $D$ mesons indicate the partial restoration of the chiral symmetry.

It is interesting to see that the behaviors of the $D$ meson masses are somewhat different from the light vector mesons such as $\rho$, $\omega$, and $\phi$ in nuclear matter. 
Their masses were predicted to decrease because of the chiral symmetry restoration in many older works (e.g., Refs.~\cite{Brown:1991kk,Hatsuda:1991ez}). 
More recent studies based on effective models, however, rather point towards a combination of strong broadening and a negative mass shift (e.g., Refs.~\cite{Rapp:1995zy,Klingl:1997kf,Leupold:1997dg}). 
On the other hand, the mass enhancement for the $D$ meson may be understood as a shift towards the degeneracy of the chiral partners (or parity partner), namely pseudoscalar $D$ and scalar $D_0$ mesons.
From this point of view, one would expect the $D_0$ meson mass to decrease with increasing density. 
This expectation is consistent with what one obtains in the OPE, in which the signs of the Wilson coefficients in front of the 
chiral-symmetry-broken condensates, $\langle \bar{q} q \rangle$ and $ \langle \bar{q} g \sigma G q \rangle$, 
for the $D$ meson channel [Eq.~(\ref{OPE_vacuum})] are opposite to those for the $D_0$ meson. 
For the light vector mesons such as $\rho$ and $\omega$, these terms 
also have a different sign from the $D$ meson, but are 
suppressed by the light quark mass and do not give a significant contribution. Other terms are more important 
for these channels and their mass shifts are therefore of somewhat different origin. 

Our results qualitatively agree with Ref.~\cite{Hilger:2008jg}, where the Borel sum rule with a ``pole + continuum'' ansatz 
were employed and a mass shift of $+45 \mathrm{MeV}$ at nuclear saturation density $\rho_0$ was obtained for 
the average of $D^+$ and $D^-$. 
Moreover, the degeneracy between the heavy-light chiral partners ($D$-$D_0$) near the critical temperature (or density) is discussed from the point of view of effective models \cite{Roder:2003uz,Blaschke:2011yv,Sasaki:2014asa,Suenaga:2014sga}.

{\it Discussion 2: $D^+$--$D^-$ mass splitting}.
In Fig.~\ref{mass shift}, we see that the $D^-$ meson mass shift (about $+38 \mathrm{MeV}$ at $\rho_0$) is stronger than that of the $D^+$ meson (about $+23 \mathrm{MeV}$ at $\rho_0$).
The mass splitting (defined $m_{D^+} - m_{D^-}$) between the $D^+$ and $D^-$ mesons is thus about $-15 \mathrm{MeV}$ at $\rho_0$.
From the viewpoint of QCD sum rule, the $D^+$--$D^-$ mass splitting is caused by the sign of the charge-symmetry-breaking $q_0$-odd terms, $ \langle q^\dag q \rangle$, $\langle q^\dag \overrightarrow{ D }_0^2 q \rangle$, and $\langle q^\dag g \sigma G q \rangle$ in Eq.~(\ref{OPE}).
We note that in Ref.~\cite{Hilger:2008jg}, a mass splitting of $-60 \mathrm{MeV}$ was obtained.

This behavior can be understood by the following intuitive physical pictures.
The $D^-$ meson has one light ``quark" which repulsively interacts with the quarks in the nuclear medium from Pauli blocking.
As a result, the bound state is weakened and the meson mass increases.
On the other hand, the $D^+$ meson has one light ``anti-quark" instead of one quark, so that it should be not affected by the Pauli blocking between quarks.
As an alternative picture, we mention the scalar and vector meson mean fields as pointed out in Refs.~\cite{Tsushima:1998ru,Sibirtsev:1999js}.
The contribution from the scalar (vector) mean field has the same (opposite) sign between a light quark and a light antiquark.
As a result, the vector mean field induces the $D^+$--$D^-$ mass splitting.
These are, however, just intuitive pictures, and in reality we have to take into account also other effects for a full understanding.

{\it Other discussions}.
Let us here mention the potential effect of in-medium broadening of the $D$-meson peaks, which was discussed in works based on hadronic effective theories.
The width broadening may be attributed to some physical origins such as resonant-hole excitations $Y_c N^{-1}$ of a charmed baryon $Y_c$ and a nucleon hole $N^{-1}$ \cite{Lutz:2005vx,Mizutani:2006vq,Tolos:2007vh,Tolos:2009nn,JimenezTejero:2011fc}. 
As sum rules, however, only provide integrals of the spectral function, they are generally not very sensitive to peak widths as long as the width is much smaller than the mass.
This is reflected in our MEM analysis, which has only a limited resolution and cannot extract detailed structures of the spectral function.
This can be understood, for instance, from our vacuum spectral function shown in Fig.~\ref{SPFvacuum}, in which the relatively large width of the $D$-meson peak can only be an MEM artifact and has no physical meaning.
Furthermore, it is seen in Fig.~\ref{SPF_DM} that the width moreover depends on the default model.
Therefore, we can in this study not make any claim about the broadening of the $D$-meson line shapes.
We emphasize, however, that in contrast to the peak width, the position of the peak can correctly be extracted from the MEM, even if its width broadens physically \cite{Gubler:2014pta}.

Finally, we comment on the error regions in Fig.~\ref{mass shift}, which come from the uncertainties of the in-medium condensates. 
The main source of this error is the $\pi N$ sigma term. 
To get a better idea on the precision of our analysis, we will check the sigma term dependence of the mass shifts in the next subsection. 

\subsection{Sigma term dependence of medium modification}

The $\pi N$ sigma term is defined as the nucleon matrix element $\sigma_{\pi N} = m_q \langle N| (\bar{u}u + \bar{d} d) |N \rangle$, with 
$m_q = (m_u + m_d)/2$. 
As mentioned earlier, 
this is a parameter related to the density dependence of the chiral condensate as 
$\langle \bar{q} q \rangle_{\rho} = \langle \bar{q} q \rangle_{0} + \sigma_{\pi N} \rho /(2m_q)$. 
The still most commonly used value of $\sigma_{\pi N}$, obtained from a phenomenological estimation \cite{Gasser:1990ce}, is $45 \mathrm{MeV}$, which we employed in the previous subsection.
The values reported in recent lattice QCD and more phenomenological studies are unfortunately still not consistent and lie roughly in the range $30-75 \mathrm{MeV}$ \cite{Ohki:2008ff,Young:2009zb,Ishikawa:2009vc,Alexandrou:2009qu,Durr:2011mp,Horsley:2011wr,Shanahan:2012wh,Bali:2012qs,Alarcon:2011zs,Hoferichter:2015dsa,Durr:2015dna}. 

\begin{figure}[tbh!]   
    \begin{minipage}[t]{1.0\columnwidth}
        \begin{center}
            \includegraphics[clip, width=1.0\columnwidth]{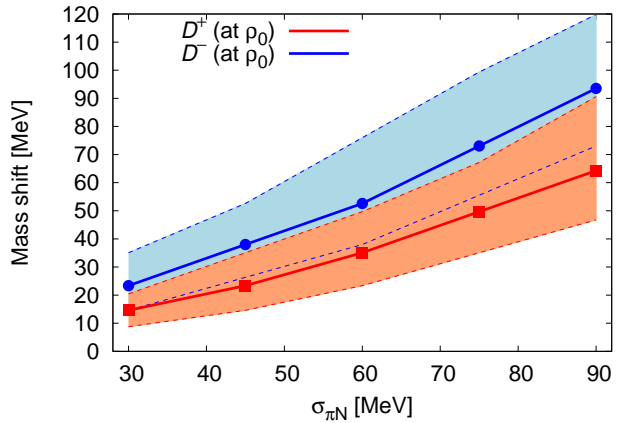}
        \end{center}
    \end{minipage}
    \caption{Sigma term dependence of $D$ meson mass shifts at nuclear saturation density $\rho_0$. 
Dashed lines and shaded areas correspond to errors from uncertainties of in-medium condensates {\it excluding the error of $\sigma_{\pi N}$}.}
    \label{sigma dependence}
\end{figure}

In this subsection, we therefore investigate the response of different sigma term values to our sum rules. 
Sigma term dependencies of the $D$ meson mass shifts at nuclear saturation density are shown in Fig.~\ref{sigma dependence}.
With a larger sigma term, both $D^+$ and $D^-$ masses exhibit increasing positive mass shifts. 
This behavior comes from the density dependence of $m_c \langle \bar{q} q \rangle$ which is proportional to the sigma term.  
On the other hand, the $D^+$--$D^-$ mass splitting expectedly shows almost no $\sigma_{\pi N}$ dependence, as the 
splitting is caused by the $q_0$-odd terms, which are not directly related to $\sigma_{\pi N}$. 
Furthermore, it becomes clear from Fig.~\ref{sigma dependence} that the density dependence of the chiral 
condensate indeed is responsible for a large part of the $D$ meson mass shift. 

Thus, the behavior of the $D$ meson in nuclear matter is quite sensitive to value of $\sigma_{\pi N}$, so that a precise evaluation of the sigma term 
will be needed to constrain the error regions shown in Fig.~\ref{mass shift}. 

\subsection{Heavy quark mass dependence of medium modification} \label{Subsection_mh_dep}

Next, we examine behaviors of the spectra when the heavy quark mass is artificially changed.
In QCD sum rules, what we need to do is only to replace the charm quark mass with an arbitrary heavy quark mass.
Strictly speaking, we have to take into account the running coupling constant $\alpha_s$ which depends on the quark flavor and the renormalization point.
We, however, here keep the coupling constant fixed because our purpose is to investigate the dependences by the heavy quark mass.

\begin{figure}[t!]
    \begin{minipage}[t]{1.0\columnwidth}
        \begin{center}
            \includegraphics[clip, width=1.0\columnwidth]{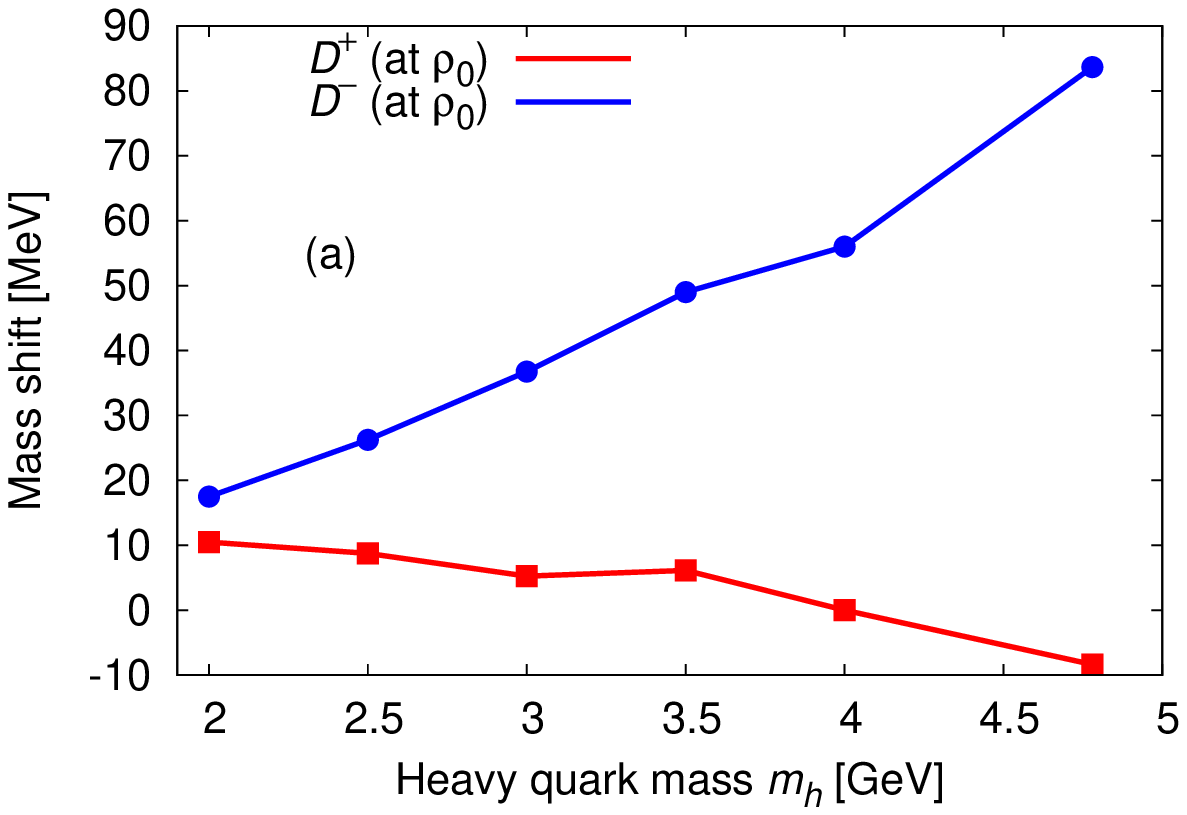}
        \end{center}
    \end{minipage}
    \begin{minipage}[t]{1.0\columnwidth}
        \begin{center}
            \includegraphics[clip, width=1.0\columnwidth]{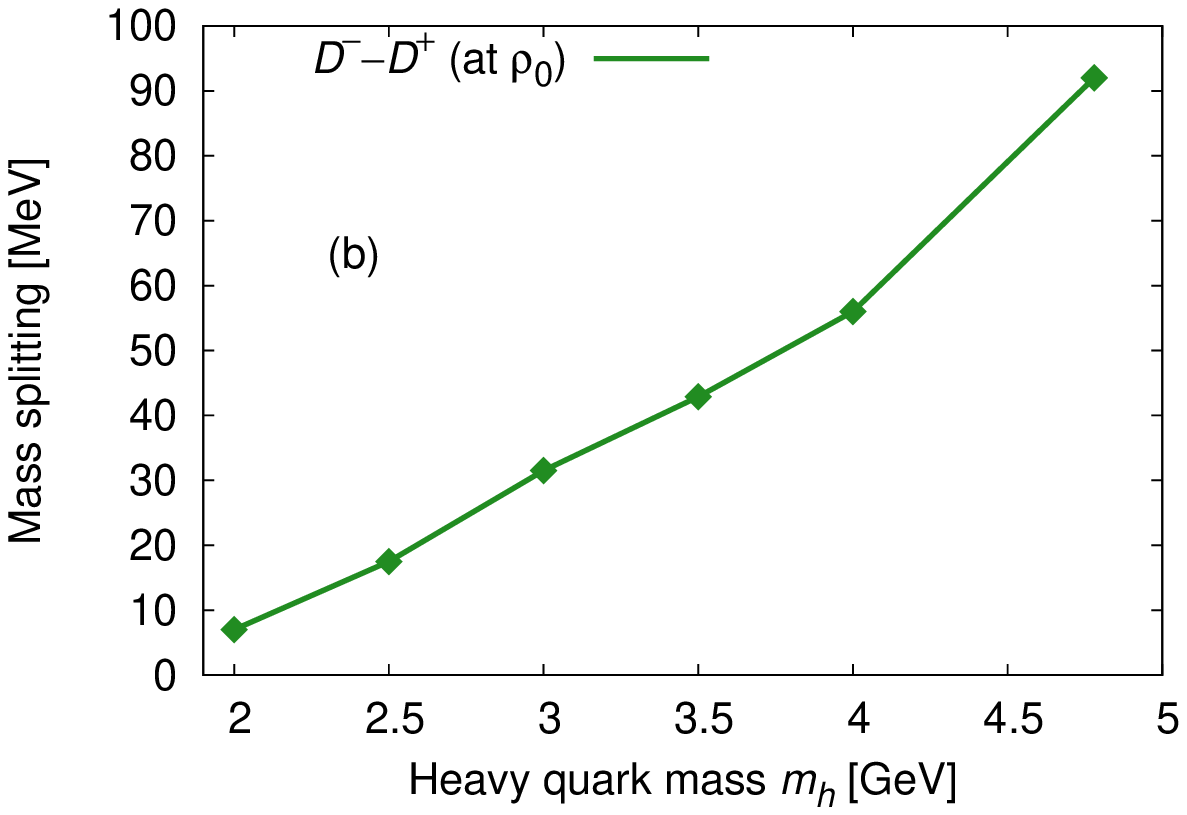}
        \end{center}
    \end{minipage}
    \caption{Artificial heavy quark mass dependence of $D$ meson mass shifts at nuclear saturation density $\rho_0$. 
(a) Dependencies of individual $D^+$ and $D^-$ mass shifts. 
(b) Dependence of mass splitting between $D^+$ and $D^-$, namely the values of $m^-(m_h, \rho=\rho_0) - m^+(m_h, \rho=\rho_0)$.}
    \label{heavy quark mass dependence}
\end{figure}

Figure~\ref{heavy quark mass dependence} shows the mass shifts of $D^+$ and $D^-$ at $\rho=\rho_0$ as a function of 
the heavy quark pole mass $m_h$, ranging from $m_h=2.0\,\mathrm{GeV}$ to the bottom quark mass $m_b=4.78\,\mathrm{GeV}$. 
In the top panel of this figure, we observe an enhanced $D^-$-meson mass with increasing the heavy quark mass.
This is caused by the heavy quark mass factor $m_h$ in the Wilson coefficient of the chiral condensate $\langle \bar{q} q \rangle$ 
and the $\langle q^{\dagger} q \rangle$ term, which have the same sign for $D^-$ and both have the effect of enhancing the 
positive mass shift. 
On the other hand, the $D^+$ mass seems to decrease for higher heavy quark masses. 
This difference can be explained as follows: 
Because the OPE of  $D^+$ and $D^-$ have a different sign in front of the $q_0$-odd terms, 
the $\langle q^{\dagger} q \rangle$ term will for the $D^+$ meson suppress the effect of the chiral condensate 
and even lead to a negative mass shift. 
To summarize, the $D^-$-meson mass is enhanced by the combination of (the heavy quark mass dependence of) the $q_0$-even and $q_0$-odd terms, 
while the reduction in the $D^+$ meson mass means that the $q_0$-odd terms overcome the $q_0$-even terms. 

These results are qualitatively consistent with the $B$-meson analysis in \cite{Hilger:2008jg}, where a mass shift of $+60\,\mathrm{MeV}$ for 
the central value of the $B^+$ and $B^-$ masses and a mass splitting of $+130\,\mathrm{MeV}$ was obtained at $\rho_0$.

\subsection{Contribution of dimension--6 condensates} \label{Subsection_dim6}
In the above analyses, condensates up to dimension--5 have so far been included. 
In this subsection, we will investigate the potential influence of dimension--6 condensates on our results. 
In Ref.~\cite{Buchheim:2014rpa}, Wilson coefficients of a large number of dimension--6 condensates in medium were 
computed for the pseudoscalar $D$ meson channel. 
The OPE provided in \cite{Buchheim:2014rpa} contains in total 14 different operators, $\langle O_1 \rangle ... \langle O_{14} \rangle$. 
At present, it is beyond our ability to give reliable estimates for the expectation values of all of these operators. 
Our analysis should therefore not be considered to be complete and final, but is rather a first order-of-magnitude estimation 
of the numerical magnitudes of these terms. 
Specifically, we will consider only five operators, namely, $\langle O_1 \rangle$, $\langle O_2 \rangle$, $\langle O_8 \rangle$, $\langle O_9 \rangle$ 
and $\langle O_{14} \rangle$, which are defined as 
\begin{eqnarray}
&& O_1 = \bar{q} \gamma^\nu t^A q \sum_f \bar{q}_f \gamma_\nu t^A q_f, \ \ \ O_2 = \bar{q} \Slash{v} t^A q \sum_f \bar{q}_f \Slash{v} t^A q_f,\nonumber\\ 
&& O_8 = \bar{q} (i v \cdot \overleftarrow{ D })^3 \Slash{v} q, \ \ \ \ \ \ \ \ \ \ \ \ \ \ \ \ O_9 = \bar{q} t^A q \sum_f \bar{q}_f \Slash{v} t^A q_f, \nonumber\\
&& O_{14} = \bar{q} (i v \cdot \overleftarrow{ D })^3 q. 
\label{dimension6condensates}
\end{eqnarray}
Their Gaussian-transformed Wilson coefficients are summarized in Appendix~\ref{App_dimension6}.

For the four-quark condensates, $\langle O_1 \rangle$, $\langle O_2 \rangle$, and $\langle O_9 \rangle$, precise 
evaluations are presently still not feasible. 
One method, that at least provides a crude estimate of these values, is the 
factorization hypothesis, in which the 
four-quark condensates are factorized into two-quark condensates.
In the linear density approximation this leads 
to \cite{Buchheim:2014rpa}: $\langle O_1 \rangle \approx -\frac{4}{9} \kappa_1 \left[ \langle \bar{q} q \rangle_0^2 (1 - 2\sigma_{\pi N} \rho/m_\pi^2f_\pi^2 ) \right]$, $\langle O_2 \rangle \approx -\frac{1}{9} \kappa_2 \left[ \langle \bar{q} q \rangle_0^2 (1 - 2\sigma_{\pi N} \rho/m_\pi^2f_\pi^2 ) \right]$,
and $\langle O_9 \rangle \approx -\frac{4}{3} \kappa_3 \langle \bar{q} q \rangle_0 \rho$, with $\kappa_1= \kappa_2/3 = \kappa_3 =1$. 
Furthermore, the traceless parts of the dimension--6 derivative condensates, $\langle O_8 \rangle$ and $\langle O_{14} \rangle$, can be estimated by the third moments 
of the quark parton distribution function ($A^q_4$) and the twist--3 parton distribution function ($e^q_3$), respectively. The results read, $\langle O_8 \rangle - \langle O_8 \rangle_\mathrm{scalar} = \langle O_8 \rangle - \frac{1}{48} g^2 \langle O_1 \rangle \approx -\frac{5}{32}m_N^3 A_4^q$ and $\langle O_{14} \rangle \approx -\frac{1}{2}m_N^3 e_3^q$, where $A_4^q =0.066 \pm 0.007$ \cite{Martin:2009iq} and $e_3^q = (1.4\pm 7.5) \times 10^{-3}$ \cite{Courtoy:2014ixa} are extracted in the same way as the second moments explained earlier. 
Note that the above expressions ignore potential spin-2 and spin-1 contributions to $\langle O_8 \rangle$ and $\langle O_{14} \rangle$. 

Adding these condensates to our sum rules, we extracted the $D$ meson spectral functions at nuclear saturation density 
and compared them to the ones obtained in the previous sections. 
As a result, we found that the dimension--6 terms give no relevant contribution to the $D$ meson mass shift in medium. 
The curves in Figs.~\ref{SPFdensity} and \ref{mass shift} indeed look identical with and 
without these terms being taken into account. 
Therefore, we can conclude that our results are not likely to depend much on the condensates shown in Eq.~(\ref{dimension6condensates}). 
To reach a definitive conclusion, a full analysis of all possible dimension--6 condensates will, however, be needed. 

\section{comments for previous works in QCD sum rules} \label{Section_Comments}
In this section, we compare our results with those of previous works, which are summarized in Table~\ref{D_meson_in_nuclear_matter_list}.

\begin{table*}[tb!]
  \begin{center}
  \begin{tabular}{l|cc|c}
\hline \hline
                                              & $\delta_{D^+}$[MeV] & $\delta_{D^-}$[MeV] & Ref. \\
\hline
Coupled channel approach (for flavor $SU(3)$) & $(*)$         &                 & \cite{Tolos:2004yg} \\
(for flavor $SU(4)$)                          & $-32$         & $+18$           & \cite{Lutz:2005vx} \\
                                              & $-(12$-$18)$  & $+(11$-$20)$    & \cite{Mizutani:2006vq,Tolos:2007vh} \\
                                              & $-35$         & $+(27$-$35)$    & \cite{JimenezTejero:2011fc} \\
(for spin-flavor $SU(8)$)                     &               & $\simeq -(20$-$27)$ & \cite{Tolos:2009nn} \\
\hline
Chiral model                                  &               & $\simeq -(30$-$180)$ & \cite{Mishra:2003se} \\
                                              & $-81$         & $-30$           & \cite{Mishra:2008cd} \\
                                              & $-77$         & $-27$           & \cite{Kumar:2010gb,Kumar:2011ff} \\
Pion exchange model                           &               & $-35.1$         & \cite{Yasui:2012rw} \\
\hline
Quark-meson coupling (QMC) model              & \multicolumn{2}{c|}{$-60$}      & \cite{Tsushima:1998ru}\\
                                              & $\simeq -140$ & $\simeq +20$    & \cite{Sibirtsev:1999js}\\
\hline
QCD sum rule $(**)$                           & \multicolumn{2}{c|}{$-48 \pm 8$}& \cite{Hayashigaki:2000es}\\
                                              & $+15$         & $+75$           & \cite{Hilger:2008jg}\\
                                              & \multicolumn{2}{c|}{$-46 \pm 7$}& \cite{Azizi:2014bba}\\
                                              & \multicolumn{2}{c|}{$-72 \pm 14\pm 9$}& \cite{Wang:2015uya}\\
                                              & $+23$         & $+38$          & This work \\
\hline \hline
  \end{tabular}
  \end{center}
   \caption{List of $D^+$- and $D^-$- meson mass shifts in nuclear medium at nuclear saturation density $\rho_0$ from various theoretical approaches. $(*)$: Reference~\cite{Tolos:2004yg} observed the quasiparticle $D^+$ peak to mix with a
resonance structure in nuclear medium.
$(**)$: References~\cite{Hayashigaki:2000es,Azizi:2014bba,Wang:2015uya} evaluated only the average mass shift of $D^+$ and $D^-$.
Reference~\cite{Hilger:2008jg} obtained the average mass shift of $+45 \mathrm{MeV}$ and $D^+$--$D^-$ mass splitting of $-60 \mathrm{MeV}$, from which we estimate the individual values of $D^+$ and $D^-$.} 
   \label{D_meson_in_nuclear_matter_list}
\end{table*}

We in particular will comment and shortly discuss 
the results of the in-medium $D$ meson masses from QCD sum rules \cite{Hayashigaki:2000es,Hilger:2008jg,Azizi:2014bba,Wang:2015uya}.
In Ref.~\cite{Hayashigaki:2000es}, the OPE with in medium condensates up to dimension--4 was used.
However, as pointed out in Ref.~\cite{Zschocke:2011aa}, one Wilson coefficient in \cite{Hayashigaki:2000es} was not correct, causing an erroneous minimum in the 
lower region of the Borel curve of the $D$ meson mass in vacuum. 
The condensates up to dimension--5 and $q_0$-odd terms were included in Ref.~\cite{Hilger:2008jg}.
Additionally, contributions from the dimension--6 four quark condensates in medium were estimated in Ref.~\cite{Buchheim:2014rpa}.
$\mathcal{O}(\alpha_s)$ corrections of the chiral condensate, $m_c \alpha_s \langle \bar{q} q \rangle$ term were calculated in Ref.~\cite{Wang:2015uya}. 

As shown in Table~\ref{D_meson_in_nuclear_matter_list}, the sign of the resulting mass shifts in Refs.~\cite{Hayashigaki:2000es,Azizi:2014bba,Wang:2015uya} is opposite 
to that in Ref.~\cite{Hilger:2008jg}. 
It should be emphasized that the 
main reason for this discrepancy lies in the difference between the choices of the Borel windows.
The approaches of Refs.~\cite{Hayashigaki:2000es,Azizi:2014bba,Wang:2015uya} relate the spectral function to the forward $D$--$N$ scattering amplitude in the 
limit of vanishing three-momentum \cite{Koike:1996ga}.
In this method, the vacuum and in-medium parts of the correlation function in nuclear medium are completely separated, so that one can focus only on the in-medium part.
As a result, the window in Ref.~\cite{Hayashigaki:2000es} is located in the region $1.73\,\mathrm{GeV}<M< 2.83 \,\mathrm{GeV}$ on the Borel curve of the mass shift. 
The Borel windows in Refs.~\cite{Azizi:2014bba} and \cite{Wang:2015uya} correspond to $2.00\,\mathrm{GeV}<M<2.83 \,\mathrm{GeV}$ 
and $2.10\,\mathrm{GeV}<M<2.32\, \mathrm{GeV}$, respectively.
On the other hand, in Ref.~\cite{Hilger:2008jg}, the Borel window was determined as $0.86\,\mathrm{GeV}<M< 1.14 \,\mathrm{GeV}$, which is clearly lower than in Refs.~\cite{Hayashigaki:2000es,Azizi:2014bba,Wang:2015uya}.
The Borel mass $M$ enters the sum rules as a factor $e^{-\omega^2/M^2}$, which strongly suppresses spectral contributions to the sum rules that lie at energies much above $M$.
This means that only if $M$ is chosen small enough, the ground state will dominate the sum rule.
Conversely, if $M$ is too large, excited states and various continuum channels will contribute to the sum rules with the comparable weight to the ground state and therefore will contaminate the result. 
This is why one usually demands that the so-called ``pole contribution" should be above $50\%$ when defining the Borel window.
This criterion is not fulfilled for the large Borel masses used in Refs.~\cite{Hayashigaki:2000es,Azizi:2014bba,Wang:2015uya} (see, for instance, the discussion given in Sec.~III of Ref.~\cite{Wang:2015uya}).
It is therefore plausible that the modifications of the excited states and the continuum 
channels at finite density are the reason for the negative mass shifts of Refs.~\cite{Hayashigaki:2000es,Azizi:2014bba,Wang:2015uya}.
This interpretation is consistent with the behavior of the mass shift Borel curves of Refs.~\cite{Hayashigaki:2000es,Wang:2015uya}, which indeed approach zero when the Borel mass is lowered towards $M \sim 1.5 \mathrm{GeV}$, showing that once the excited states are removed from the sum rules, the claimed negative mass shift in fact vanishes (in Ref.~\cite{Azizi:2014bba} the Borel curve is not shown for such small Borel masses).
We therefore, believe that the results obtained from the smaller Borel masses of Ref.~\cite{Hilger:2008jg} are more 
reliable.

Our window used as an input into MEM is compatible with that of Ref.~\cite{Hilger:2008jg}.
Here, we stress that our results for the mass of the ground state do not depend on the threshold parameter or the density dependence of the continuum, so that we can 
focus only on the medium modification of the ground state peak.
With the higher Borel window used in Refs.~\cite{Hayashigaki:2000es,Azizi:2014bba,Wang:2015uya}, we cannot reproduce the $D$ meson peak in vacuum from MEM because 
of the dominant continuum contribution to the sum rule. 

\section{Conclusion and Outlook} \label{Section_Conclusion and Outlook}
We have investigated the pseudoscalar $D$ meson mass in nuclear medium by using QCD sum rules and MEM.
To separate $D^+$ and $D^-$ into independent contributions, we have constructed the charge-conjugate-projected sum rules.
From these sum rules and MEM, we have obtained the spectral functions for the $D^+$ and $D^-$ mesons in nuclear matter. 
It is found that both $D^+$ and $D^-$ peaks are shifted to a higher energy with increasing density.
This result indicates the enhancement of $D$ meson mass from the partial restoration of chiral symmetry.
We have moreover observed a $D^+$--$D^-$ mass splitting of about $-15 \mathrm{MeV}$ at nuclear saturation density $\rho_0$. 
This behavior is attributed to the $q_0$-odd condensates, which break the charge symmetry.
The $D$ meson system is thus found to be useful to probe the chiral and charge symmetries at finite density. 

\begin{acknowledgments}
The authors gratefully thank Tetsuo Hatsuda, Su Houng Lee, Keisuke Ohtani, Wolfram Weise and Shigehiro Yasui for useful discussions.  
This work was partially supported by KAKENHI under Contract Nos.25247036.
K.S. was supported by Grant-in-Aid for JSPS Fellows from Japan Society for the Promotion of Science (JSPS) (Grant No.26-8288).
\end{acknowledgments}

\appendix
\section{OPE of dimension--6 condensates} \label{App_dimension6}
In this appendix, the dimension--6 part of the OPE used in Sec.~\ref{Subsection_dim6} is briefly summarized. 
In Ref.~\cite{Buchheim:2014rpa}, the Wilson coefficients of in medium 
dimension--6 condensates were computed for in total 14 different operators. 
In this work, we focus on only five of them and neglect the nine condensates which contain a gluon field. 
From Eq. (8) in Ref.~\cite{Buchheim:2014rpa}, by setting $v=(1,0,0,0)$ and $p=(1,0,0,0)$, the OPE in momentum space can be written as
\begin{eqnarray}
&& \Pi_\mathrm{dim 6}^\mathrm{even} (q_0) = \nonumber \\
&&\frac{1}{3} \frac{1}{(q_0^2-m_h^2)^2} \left[ 1+\frac{1}{2}\frac{m_h^2}{q_0^2-m_h^2} -\frac{1}{2} \frac{m_h^4}{(q_0^2-m_c^2)^2} \right] g^2 \langle O_1 \rangle \nonumber\\
\nonumber\\
&& -\frac{1}{3} \frac{q_0^2}{(q_0^2-m_h^2)^3} \left[ -\frac{9}{2} g^2 \langle O_1 \rangle + 8g^2 \langle O_2 \rangle \right] \nonumber\\
&& +\frac{1}{6} \frac{q_0^4}{(q_0^2-m_h^2)^4} \left[ g^2 \langle O_1 \rangle - 48 \langle O_8 \rangle \right], \\
&& \Pi_\mathrm{dim 6}^\mathrm{odd} (q_0) = - 2 m_h \frac{1}{(q_0^2-m_h^2)^3} g^2 \langle O_9 \rangle \nonumber\\
&& + 8 m_h \frac{q_0^2}{(q_0^2-m_h^2)^4} \langle O_{14} \rangle.
\end{eqnarray}
Furthermore, performing the Gaussian transformation of Eq.~(\ref{Gauss_trans}), we finally obtain
\begin{widetext}
\begin{eqnarray}
\tilde{G}^\mathrm{even}_{\langle O_1 \rangle} (\hat{s}, \tau) &=& g^2 \langle O_1 \rangle \frac{1}{2\sqrt{4\pi \tau}} e^{-\frac{(m_h^2 - \hat{s})^2}{4\tau}} \frac{1}{288 \tau^3} \left[ -m_h^{10} +3 m_h^8 \hat{s} -3 m_h^6 \hat{s}^2 + m_h^4  (\hat{s}^3 + 6\hat{s} \tau ) -6 m_h^2 (\hat{s}^2 -10 \tau)\tau -48\hat{s} \tau^2 \right], \nonumber\\
\\
\tilde{G}^\mathrm{even}_{\langle O_2 \rangle} (\hat{s}, \tau) &=& \left[ -\frac{9}{2} g^2 \langle O_1 \rangle + 8g^2 \langle O_2 \rangle \right] \frac{1}{2\sqrt{4\pi \tau}} e^{-\frac{(m_h^2 - \hat{s})^2}{4\tau}} \frac{1}{24 \tau^2} \left[ m_h^6 - 2m_h^4 \hat{s} + m_h^2 (\hat{s}^2 - 6 \tau ) + 4 \hat{s} \tau \right], \\
\tilde{G}^\mathrm{even}_{\langle O_8 \rangle} (\hat{s}, \tau) &=& \left[ g^2 \langle O_1 \rangle - 48 \langle O_8 \rangle \right] \frac{1}{2\sqrt{4\pi \tau}} e^{-\frac{(m_h^2 - \hat{s})^2}{4\tau}} \nonumber\\
&& \times \frac{1}{288 \tau^3} \left[ m_h^{10} - 3 m_h^8 \hat{s} + 3 m_h^6 (\hat{s}^2 - 6\tau) - m_h^4  (\hat{s}^3 - 30 \hat{s} \tau) - 12 m_h^2 (\hat{s}^2 -4 \tau) \tau -24 \hat{s} \tau^2 \right], \\
\tilde{G}^\mathrm{odd}_{\langle O_9 \rangle} (\hat{s}, \tau) &=& g^2 \langle O_9 \rangle \frac{1}{2\sqrt{4\pi \tau}} e^{-\frac{(m_h^2 - \hat{s})^2}{4\tau}} \frac{1}{4 m_h^2 \tau^2} \left[ m_h^8 - 2 m_h^6 \hat{s} + m_h^4 (\hat{s}^2 - 4\tau) + 2m_h^2 \hat{s} \tau - \tau^2 \right], \\
\tilde{G}^\mathrm{odd}_{\langle O_{14} \rangle} (\hat{s}, \tau) &=& \langle O_{14} \rangle \frac{1}{2\sqrt{4\pi \tau}} e^{-\frac{(m_h^2 - \hat{s})^2}{4\tau}} \nonumber\\
&& \times \frac{1}{6 m_h^2 \tau^3} \left[ m_h^{12} - 3m_h^{10} \hat{s} + 3 m_h^8(\hat{s}^2 - 5\tau) - m_h^6 (\hat{s}^3 - 24\hat{s} \tau ) - 9 m_h^4 ( \hat{s}^2 - 3\tau) \tau - 9m_h^2\hat{s} \tau^2 + 3 \tau^3 \right].
\end{eqnarray}

\end{widetext}

\section{MEM for QCD sum rules} \label{App_MEM}
In this section, we briefly introduce the procedure of the MEM analysis for QCD sum rules.
More technical details are shown in Ref.~\cite{Gubler:2010cf}.
The MEM is based on Bayes' theorem:  
\begin{equation}
P[\rho |\tilde{G} H] =\frac{P[\tilde{G} |\rho H] P[\rho | H]}{ P [\tilde{G}|H]}, \label{Bayes_theorem}
\end{equation}
where $\rho$ and $\tilde{G}$ correspond to the spectral function and the OPE in our sum rules (Eq.~(\ref{sum rule})), respectively.
$H$ denotes prior knowledge on $\rho$ such as positivity and its asymptotic values.
$P[\rho |\tilde{G} H]$ represents the conditional probability of $\rho$ if $\tilde{G}$ and $H$ are given. 
On the right-hand side, $P[\tilde{G} |\rho H]$ and $ P[\rho | H]$ stand for the (i) {\it likelihood function} and (ii) {\it prior probability}, respectively.
$P [\tilde{G}|H]$ is only a normalization constant and does not depend on $\rho$.
To maximize $P[\rho |\tilde{G} H]$, we have to estimate $P[\tilde{G} |\rho H]$ and $ P[\rho | H]$.

(i) The likelihood function is written as
\begin{equation}
P[\tilde{G} |\rho H] = e^{-L[\rho]}, \label{likelifood_function0}\\
\end{equation}
\begin{equation}
L[\rho] = \frac{1}{2 \hat{s}^- \tau^-} \int_{\hat{s}_{\mathrm{min}}}^{\hat{s}_{\mathrm{max}}} d\hat{s} \int_{\tau_{\mathrm{min}}}^{\tau_{\mathrm{max}}} d\tau \frac{[\tilde{G} (\hat{s},\tau) - \tilde{G}_\rho(\hat{s},\tau)]^2}{\sigma^2(\hat{s},\tau)}, \label{likelifood_function}
\end{equation}
where $\hat{s}^- = \hat{s}_{\mathrm{max}}-\hat{s}_{\mathrm{min}}$ and $\tau^- = \tau_{\mathrm{max}}-\tau_{\mathrm{min}}$.
Here, $\tilde{G} (\hat{s},\tau)$ is obtained from the OPE and corresponds to the left-hand side in our sum rules, while $\tilde{G}_\rho(\hat{s},\tau)$ is defined as the right-hand one in Eq.~(\ref{sum rule}).
$\sigma(\hat{s},\tau)$ stands for the uncertainty of $\tilde{G}(\hat{s},\tau)$ (see Ref.~\cite{Gubler:2010cf}).

(ii) The prior probability is written as
\begin{equation}
P[\rho | H] = e^{\alpha S[\rho]}, \label{shannon-jaynes_entropy0}\\
\end{equation}
\begin{equation}
S[\rho] = \int_0^\infty d\omega \left[ \rho(\omega) - m(\omega) - \rho(\omega) \log \left( \frac{\rho(\omega)}{m(\omega)} \right) \right], \label{shannon-jaynes_entropy}
\end{equation}
where $S[\rho]$ is known as the {\it Shannon-Jaynes entropy} and $\alpha$ is introduced as a real positive scaling factor.
$m(\omega)$ is called the {\it default model} and determines the spectral function when there is no information from the OPE.

Using Eqs.~(\ref{likelifood_function0}) and (\ref{shannon-jaynes_entropy0}), we rewrite Eq.~(\ref{Bayes_theorem}) as
\begin{eqnarray}
P[\rho|\tilde{G} H] &\propto& P[\tilde{G}|\rho H] P[\rho|H] \nonumber\\
&=& e^{Q[\rho]}, \\
Q[\rho] &\equiv& \alpha S[\rho] - L[\rho].
\end{eqnarray}
To determine the most probable $\rho(\omega)$, we search for the maximum of the functional $Q[\rho]$ by the Bryan algorithm \cite{Bryan:1990}.

\bibliography{reference}
\end{document}